\documentclass{article}

\usepackage[final, nonatbib]{nips_2018}

\usepackage[utf8]{inputenc} 
\usepackage[T1]{fontenc}    
\usepackage{hyperref}       
\usepackage{url}            
\usepackage{booktabs}       
\usepackage{amsfonts}       
\usepackage{nicefrac}       
\usepackage{microtype}      
\usepackage[pdftex]{graphicx}
\title{A Dynamic Network and Representation Learning Approach for Quantifying Economic Growth from Satellite Imagery}
\author{
    Jiqian Dong\thanks{Equal Contribution}\space,\ \  Gopaljee Atulya$^*$, Kartikeya Bhardwaj$^*$, and Radu Marculescu \\
    Carnegie Mellon University\\
    Pittsburgh, PA 15213 \\
    \texttt{\{jiqiand, gatulya, kbhardwa, radum\}@cmu.edu}\\
}

\begin{document}

\maketitle

\begin{abstract}\vspace{-2mm}

Quantifying the improvement in human living standard, as well as the city growth in developing countries, is a challenging problem due to the lack of reliable economic data. 
Therefore, there is a fundamental need for alternate, largely unsupervised, computational methods that can estimate the economic conditions in the developing regions. To this end, we propose a new network science- and representation learning-based approach that can quantify economic indicators and visualize the growth of various regions. 
More precisely, we first create a dynamic network drawn out of high-resolution nightlight satellite images. We then demonstrate that using representation learning to mine the resulting network, our proposed approach can accurately predict spatial gross economic expenditures over large regions. Our method, which requires only nightlight images and limited survey data, can capture city-growth, as well as how people's living standard is changing; this can ultimately facilitate the decision makers' understanding of growth without heavily relying on expensive and time-consuming surveys.

\end{abstract}
\vspace{-3mm}
\section{Introduction}\vspace{-2mm}

The ongoing big data and machine learning revolution has greatly contributed to quantifying economic development. However, although massive data is available for developed countries, the developing countries still suffer from the lack of reliable economic data. This polarization also exists in the quality of data since the recorded data in developing countries has been known to be unstructured and inaccurate \cite{baddata15}, thereby making the useful data even more scarce. All these challenges have greatly hindered the accurate modeling of economic conditions in the developing world.

Knowing how people live and the rate of improvement of their living standard are fundamental components of many decision making processes such as urban development, policy effectiveness evaluation, and conservation planning, etc. 
However, the lack of sufficient high-quality data in developing countries can result in poor decisions which can in turn lead to bad development policies. 
On the other hand, apart from this low-quality economic data obtained from time- and labor-intensive surveys, the massive amount of high-resolution nightlight satellite images forms a more reliable source of data for developing countries. Therefore, new machine learning-based methods that can more effectively exploit the power of nightlight satellite images to predict economic indicators and explicitly quantify the economic growth of different regions. 

In this paper, we propose a new network science- and representation learning-based approach to explicitly capture the complex relationship between the economic growth of various regions, and also predict economic indicators over large regions. To this end, we propose a novel \textit{gravity-based model} to create, for the very first time, a \textit{dynamic economic-growth network} by using massive amount of nightlight satellite data. We also demonstrate that mining this network with representation learning techniques like node2vec~\cite{node2vec-kdd2016} can help us accurately predict economic indicators over large regions. 

Overall, we demonstrate that by using our proposed gravity model and publicly available satellite data, we can create large-scale economic-growth networks in a largely unsupervised manner (with the exception of fine-tuning a few hyper-parameters). Mining this network can give us features relevant to economic growth which can be further used for  accurately predicting economic indicators. This way, we can clearly obtain socio-economic features for large regions without relying on expensive and time consuming surveys. Hence, our method can be used for large-scale economic modeling in the developing world.
\vspace{-2mm}

\section{Previous Work}\vspace{-2mm}
A prior deep learning approach on nightlight satellite images for poverty prediction was proposed in \cite{Jean790}. Since using a deep learning model directly with nightlight images  requires a large data-set with millions of samples with labelled economic data, Jean et al. address this problem by applying a transfer learning technique with an ImageNet-pretrained deep network. As a result, feature embeddings from the CNN model can explain $75\%$ of variation in local-level economic outcomes \cite{Jean790,Xie2016TransferLF}. Other prior works extend such ideas to population estimation  \cite{Robinson2017ADL} and crop yield prediction \cite{Wang:2018:DTL:3209811.3212707}. 

Although this prior art predicts poverty via satellite images, this prediction is mostly spatial in nature. Consequently, the \textit{temporal} aspects of economic growth have been largely ignored in prior works. Moreover, most CNN models have an inherent problem of being \textit{non-interpretable}. This is especially critical for applications where decisions must be made based on the machine learning model output (for instance, for urban planning, etc.).  Specifically, using interpretable machine learning models can help to actually understand the underlying factors behind the model output in order to more accurately target such factors while making decisions.

In our work, we address both limitations above. Specifically, since we build economic-growth networks from the satellite data, we can analyze their properties like community structure~\cite{newman2006modularity} to understand which locations experience growth over time; this will result in a much more interpretable model than, say, the output of a CNN. Also, while previous work lacks temporal analysis and does \textit{not} predict the future development trends, our dynamic network can be potentially used for spatio-temporal predictions (and not just spatial!). We show some preliminary results for accurate spatiotemporal predictions using our proposed approach.

\section{Proposed Approach}\vspace{-3mm}
Our approach can be divided in to three parts: network construction, network mining and regression.
\subsection{Economic Gravity-Based Network Construction Model}\vspace{-2mm}
Our gravitational model aims to predict the economic indicators. To achieve this, we divide the region of interest into smaller grids and treat each grid as a node in the network. For example, we consider economic prediction for Tanzania, and thus we divided the entire country into $50\times51$ grids where each grid has an area of approximate $524.4$km$^2$ ($22.9$km $\times 22.9$km). For each node (grid), we draw undirected edges to all the other nodes with a weight calculated with the following gravitational equation:\vspace{-1mm}
\begin{equation}
    W_{i,j} = \frac{{M_i\cdot{M_j}}}{R^P}
\end{equation}
where, $W_{i,j}$ is the weight on the edge from node $i$ to $j$, $M_{i}=$ log$(I_i+1)$ is the total grid nightlight log-intensity for node $i$, and $R$ is the normalized distance between each node (ranging from $0$ to $1$). Parameter $P$ controls the trade-off between forming links based on intensity vs distance. 

After drawing edges for all nodes, the network is fully connected with total $2550\times2550$ edges. Since there are many edges with low \textit{gravity} weight, we set a threshold $\tau$ to remove low gravity links. Finally, to prevent nodes from getting disconnected from the network, we rewire the nodes with less than $K$ links to their $K$ nearest neighbors; 
 this is similar to the $K$-$\tau$ method introduced in \cite{bhardwaj2018dimensionality}.


\subsection{Network Mining}\vspace{-2mm}

\begin{figure}[!]
\centering
\includegraphics[width=5in]{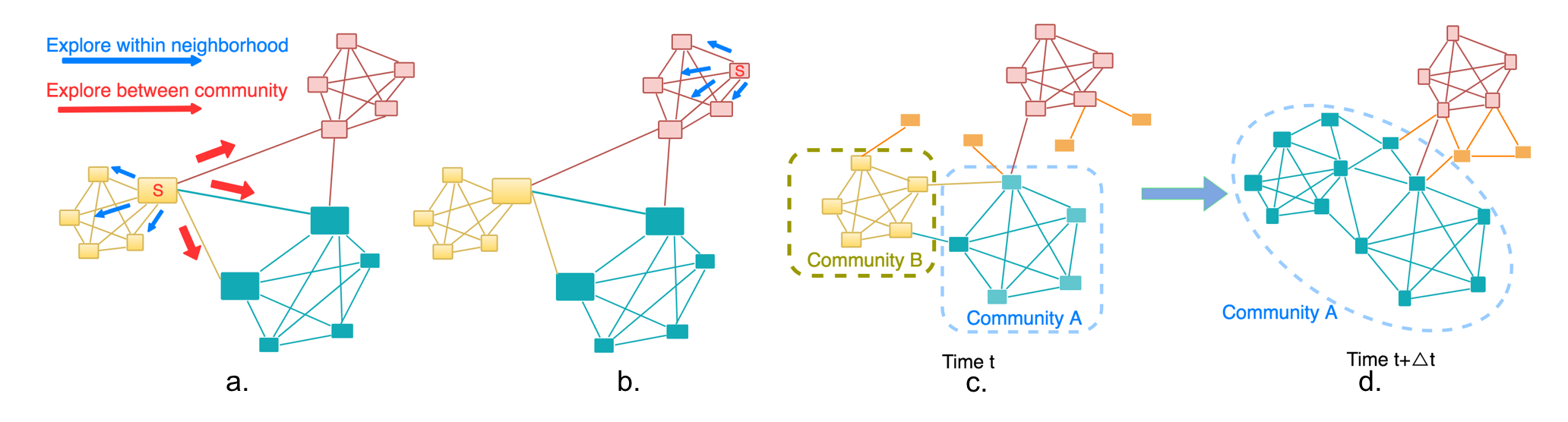}\vspace{-3mm}
	\caption{
	(a)Random walk starts from the hub node, both search the within neighborhood and between community. (b)Random work starts from the non-hub node, mainly explore the neighborhood within community. (c),(d)Dynamic network evolution: Initially, there is a community A of rich locations; nearby locations that are not as rich form a separate community B. In the next time step, locations in community B experience growth and merge into community A. node2vec on various snapshots can take these dynamically changing communities into account and generate more intuitive features.\vspace{-2mm}}
\label{dynNet}
\end{figure}

Intuitively, based on the gravity network, we can see that if a node (i.e., the location) is more developed, it will have a higher nightlight intensity and, hence, more links to nearby node; that is, if a community~\cite{newman2006modularity} of economically wealthy locations in the network grows over time, it is high likely that this region is experiencing rapid economic growth (see Fig.~\ref{dynNet}). These community-based features can be captured by existing representation learning frameworks like node2vec~\cite{node2vec-kdd2016}. The algorithm is a biased random work on a graph, and the probability to travel between two nodes is a function edge-weight between them. Two adjustable parameters, $p$ and $q$ are introduced representing priority of the walk to trade-off between exploration and exploitation. Here we choose the $q=0.5, p=1$ in node2vec implementation, smaller value of $q$ is emphasized to prefer exploitation over exploration. 

As illustrated in Fig.~\ref{dynNet} c and d, if the number of nodes in the community grows, node2vec will generate similar features for the nodes that belong to a similar community. Moreover, since the network communities share similar characteristics, these locations  will most likely share similar consumption index. Hence, the latent developing trend, as well as the similarity between nodes can be captured via network representation learning, thereby resulting in features that are easier to interpret. Also, visualizing the dynamic network can also inform decision makers of other latent trends (e.g., location X became similar to location Y in terms of economic development).

\subsection{Regression Model over Economic Gravity-Based Features}\vspace{-2mm}
We next convert the walks simulated from node2vec neighborhood search strategy into walks of light intensities by replacing the grid ID with its grid intensity. Then, the expected value of intensity at each step $i \in (1, L), L - Walk \ \ length$ originating from node $k \in (1,K), K - Number \ \ of \ \ grids$ is calculated. Therefore, the feature vector of a node represents the expected value of intensity at each step of a random walk originating from that node.

Finally, the features generated above for each node, and the corresponding economic indicators are used in a regression model (e.g., Random Forest or K-Nearest Neighbor regression or Bayesian) for training and testing. Note that, our proposed method does not require labelled economic data to build the economic-gravity network. More precisely, we can derive these community-based features for a large number of locations and use them for economic prediction on locations that do not have large-scale surveys. Therefore, our model can be used as an alternate way to quantify economic condition and growth trends in developing countries without heavily relying on surveys which are, time-, money-, and labor-intensive.

\section{Results}\vspace{-3mm}
\subsection{Experimental Setup}\vspace{-2mm}
We use monthly nightlight images from National Oceanic and Atmospheric Administration (NOAA) with a 15 arc-second spatial-resolution geographic grid \cite{noaa}. For economic prediction, we use Living Standards Measurement Study (LSMS) survey data with consumption expenditure and coordinates (measured in Longitude and Latitude) from Tanzania, National Panel Survey (NPS) for the year $2012-2013, \ \ 2014-2015$, and Malawi, Integrated Household Panel Survey (IHPS) for the year $2013, \ \ 2016$ \cite{lsms} as our ground truth for training and prediction. 
We conduct spatial and spatio-temporal predictions in the current version of this paper. For the gravity model, we set $\tau=5$ and $K=3$ for Tanzania, and $\tau=2$ and $K=1$ for Malawi as a function of the country size and density of light-intensity.

\subsection{Economic Prediction Results}\vspace{-2mm}
In this section, we present results for spatially predicting economic indicators at several locations in Tanzania and Malawi. After, preprocessing the LSMS survey data-set (cleaning, binning, and mapping) we can recover around 500-1000 data points per set per year. We split it $50\%$ for training and $50\%$ for testing. Next, we build economic-gravity network from average nightlight images available for the corresponding year. After computing the features for all the nodes in the network (which are grids on the map), we associate the each node to the nearby houses with a Manhattan distance threshold of $0.25$ degree of longitude and latitude which is equivalently an $11$km region around the grid center. We conducted experiments with $100$ different training-testing splits to obtain statistically significant results.

\begin{figure}[!]
\centering
\includegraphics[width=5.1in]{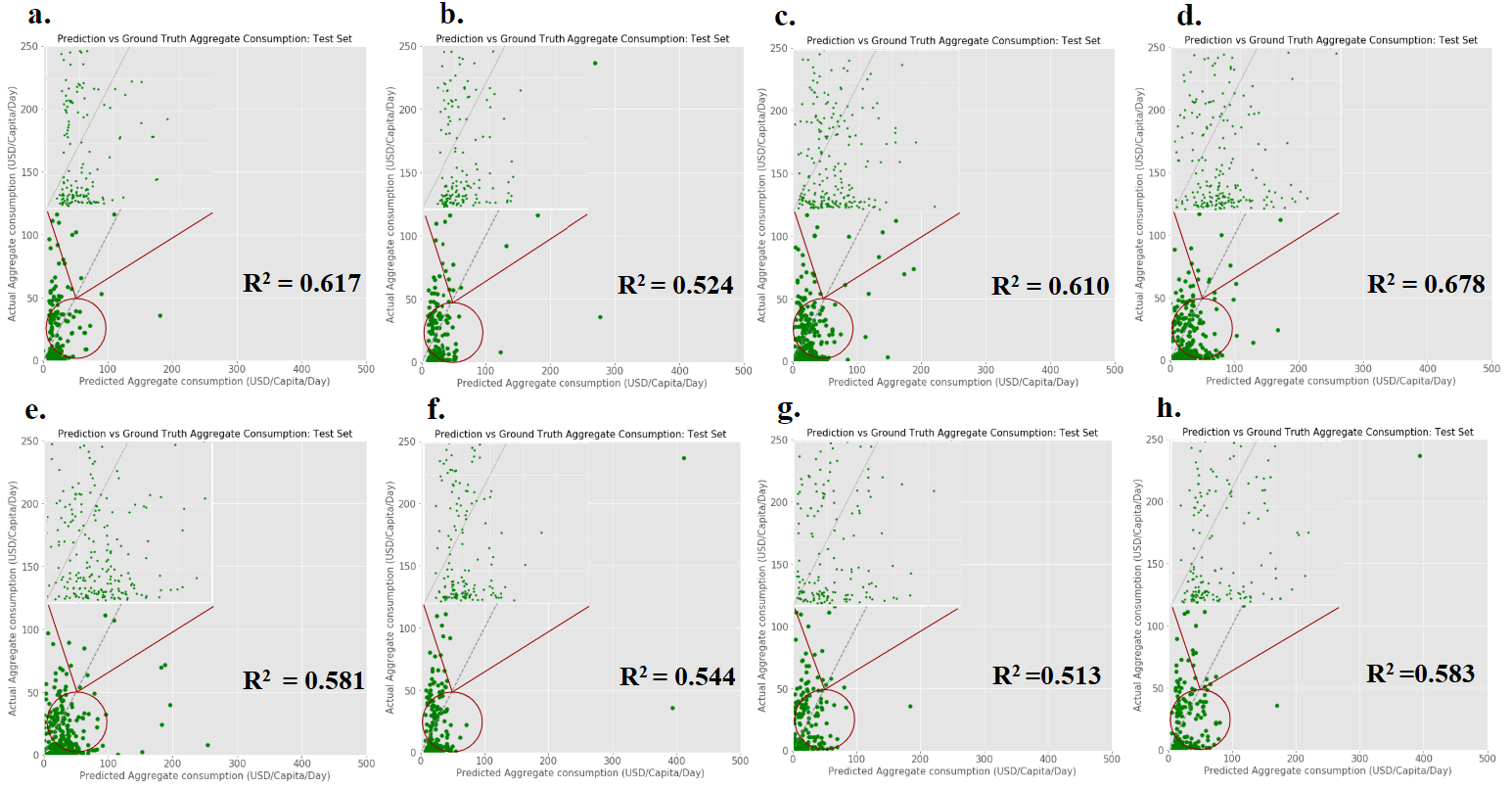}\vspace{-1mm}
	\caption{Coefficient of determination, $R^2$ on test sets for (a) Bayesian-Ridge Regression (BRR), (b) Random-Forest Regression (RFR), (c) K-Nearest-Neighbours Regression (KNNR) (d) Linear Regression (LR) for Tanzania 2013. Similarly, (e), (f), (g) and (h) represent $R^2$ on test sets for BRR, RFR, KNNR, and LR respectively for Malawi 2013. High $R^2$ for testing set shows that node2vec features for economic-gravity networks indeed capture economic information.\vspace{-1mm}}
\label{res}
\end{figure}

\begin{table}
\begin{center}
\caption{Spatial Predictions - Test $R^2$ by region and year for models BRR, RFR, KNNR, and LR.\vspace{-2mm}}
\label{tab1}
\begin{tabular}{ |c|c|c|c|c|c| } 
\hline
Country & Year & BRR Test $R^2$ & RFR Test $R^2$ & KNNR Test $R^2$ & LR Test $R^2$ \\
\hline
Tanzania & 2013 & 0.61718 & 0.52485 & 0.61048 & 0.67821\\ 
\hline
Tanzania & 2015 & 0.60326 & 0.41516 & 0.47981 & 0.63308\\ 
\hline
Malawi & 2013 & 0.58167 & 0.54436 & 0.51313 & 0.58330\\ 
\hline
Malawi & 2016 & 0.44194 & 0.45513 & 0.34388 & 0.43937\\ 
\hline
\end{tabular}
\end{center}
\vspace{-4mm}
\end{table}

\begin{figure}[!]
\centering
\includegraphics[width=4.1in]{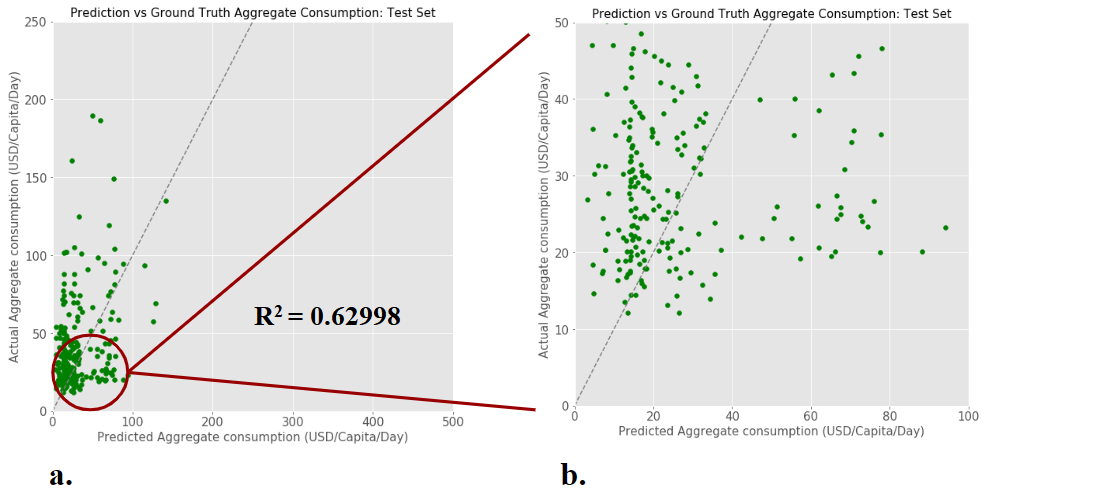}\vspace{-1mm}
	\caption{Spatio-temporal Predictions - (a) coefficient of determination, $R^2$ for aggregate consumption of Tanzania 2015 from features weights learned from Tanzania 2013. K-Nearest-Neighbours Regression Model is used for prediction. (b) is an enlarged version of (a).\vspace{-3mm}}
\label{res2}
\end{figure}

We evaluate our model via calculating $R^2$, which represents the amount of variance explained by the model. The prediction results (for the median $R^2$ value) are shown in Fig.~\ref{res}. As evident, Fig.~\ref{res}(a) shows $R^2$ for the Random Forest model used on the training set, while Fig.~\ref{res}(b) shows $R^2$ for the testing set for the same model. Similarly, Fig.~\ref{res}(c) shows $R^2$ for the testing set with the K-Nearest Neighbor model. Clearly, the $R^2$ on the testing set for both regression models is high which indicates that features generated by the node2vec algorithm on the proposed gravitational network capture very relevant economic features. The $R^2$ results for the rest of the experiments are summarized in Table~\ref{tab1}.

Fig.~\ref{res2}(a, b) show the spatio-temporal results for Tanzania 2015 when the night-light imagery data from Tanzania 2013 is used for training the models, indicating features weights learned through random walks are robust, and transferable temporally. Therefore, our model can be used to quantify large-scale economic conditions in developing countries without relying on expensive survey data. Of note, our test-set $R^2=0.5$ is close to the $R^2$ reported in the prior work for Tanzania~\cite{Jean790}.

\subsection{Growth Monitoring via Community Detection on Economic-Gravity Networks}
\begin{figure}[!]
\centering
\includegraphics[width=5in]{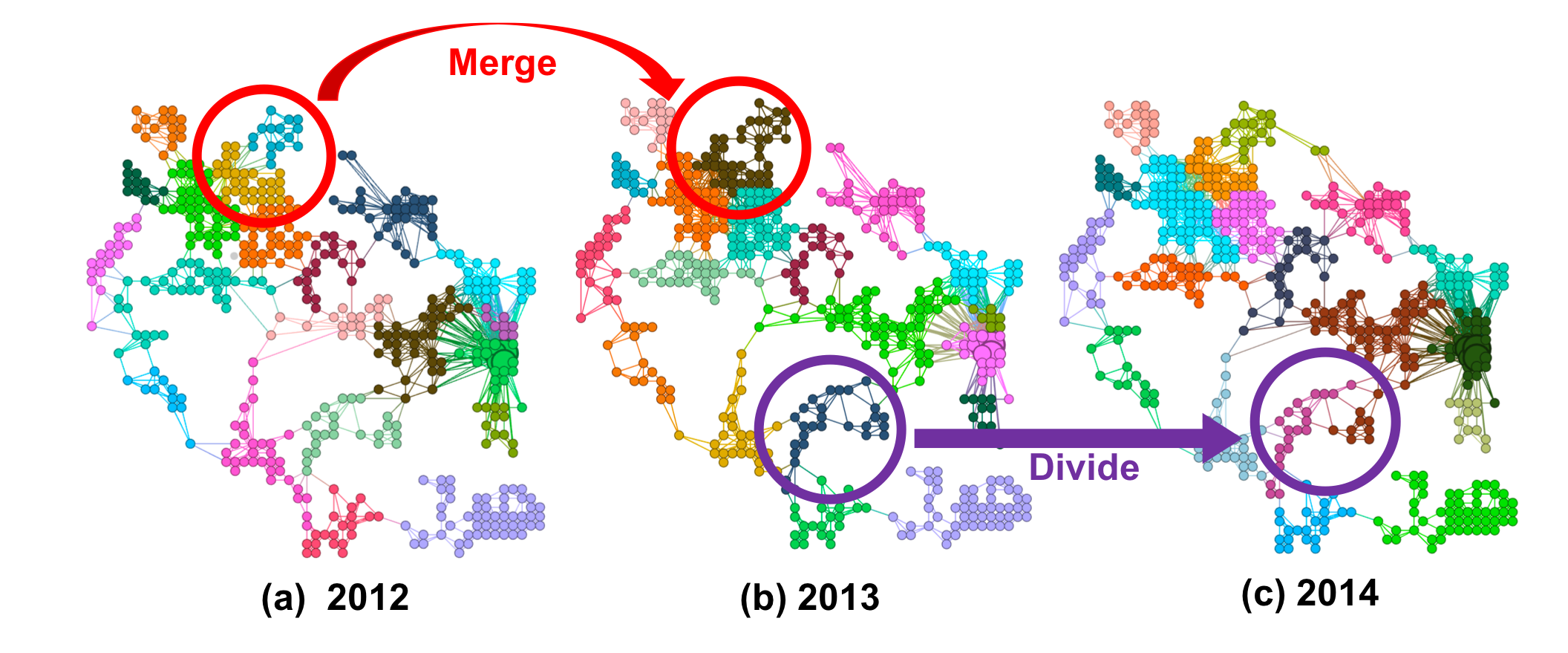}\vspace{-3mm}
	\caption{Dynamic community structure for Tanzania: (a) 2012 has 23 communities, (b) 2013 has 20, and (c) 2014 has 18 communities. This shows that the communities merge due to similar economic growth trends. Some communities also get divided possibly due to different growth trends.\vspace{-2mm}}
\label{comm}
\end{figure}

\vspace{-2mm}
For monitoring economic growth-similarity trends, we show our results in Fig.~\ref{comm}(a)-(c) for high resolution community detection on economic-gravity networks for 2012, 2013, and 2014 nightlight data.
The number of communities in the network varied from 23 in 2012, to 20 in 2013, to 18 in 2014. This is in agreement with our initial hypothesis presented in Fig.~\ref{dynNet} that when economic growth becomes similar among different regions, they will get merged into bigger communities. This is clearly evident from Fig.~\ref{comm}(a,b) when the communities merge due to similar growth trends in the Northwestern part of Tanzania (red arrows), while a community breaks down into two in the Southeastern part (violet arrow). Therefore, these dynamic growth trends can be  easily captured by our network-based approach. This further provides more information to decision makers about which locations are experiencing similar/different growth rates.\vspace{-1mm}

\section{Conclusion and future work}\vspace{-3mm}
We have proposed a new dynamic gravity-based network model for quantifying economic growth across a large region. To this end, we have used representation learning on a gravity-network to extract different growth-related features.
Our results have demonstrated that our proposed approach can indeed accurately predict the spatial and spatio-temporal gross economic expenditures.
We have further used dynamic community structure to monitor the growth of different regions over time.

For future work, we plan to improve upon our current temporal economic predictions by building representation learning methods for dynamically changing networks.

\bibliography{reference}
\bibliographystyle{unsrt}






\end{document}